
%

\input phyzzx
 \hsize=15.8cm
\vsize=23cm

\def\picture #1 by #2 (#3){
  \vbox to #2{
   \hrule width #1 height 0pt depth 0pt
    \vfill
   \special{picture #3}     }
  }

\def\scaledpicture #1 by #2 (#3 scaled #4){{
\dimen0=#1 \dimen1=#2
  \divide\dimen0 by 1000 \multiply\dimen0 by #4
  \divide\dimen1 by 1000 \multiply\dimen1 by #4
  \picture \dimen0 by \dimen1 (#3 scaled #4)}
  }

\def\FIG#1{\global\advance\figurecount by 1
     \xdef#1{\chapterlabel\number\figurecount}}

\parskip=0pt plus 1pt
\def\figure#1{\midinsert\Tenpoint\vskip#1\noindent}
\def\endfigure{\endinsert\par\vskip20pt}

\parindent 25pt
\overfullrule=0pt
\tolerance=10000
\def\ie{{\it i.e.}}

\sequentialequations
\nopagenumbers

\hfill {QMW-91-24}\break
 \null
\vskip 5cm
\centerline{TEMPERATURE DEPENDENCE OF STRING THEORY IN THE}
\centerline{PRESENCE OF WORLD-SHEET BOUNDARIES}
\vskip 1cm
 \centerline{ Michael B.  Green,}
\centerline{Department of Physics,}
\centerline{Queen Mary and Westfield College,  London, UK}

\nopagenumbers
\vskip 4cm
\abstract
The effect of world-sheet boundaries on the
temperature-dependence of bosonic string theory is studied
to first order in string perturbation theory.  The high-
temperature
behaviour of a theory with Dirichlet boundary conditions has
features
suggestive of the high-temperature limit of the confining
phase of large-$n$ $SU(n)$ Yang--Mills theory, recently
discussed by
Polchinski.
\vfill\eject
\pagenumbers

The question of whether there is an exact correspondence
between QCD and some kind of interacting string theory is of
long-standing interest.  Although the quark and gluon
picture of the structure of hadrons is simple and useful for
discussing scattering at high energy and fixed angle or deep
inelastic processes it is an awkward way of describing the
bound states of the theory.   The possibility
of expressing QCD as some kind of string theory is motivated
both by the
lattice formulation
of $SU(n)$ Yang--Mills theory as well as by the large $n$
limit in the continuum.   Holding $g^2 n $
fixed as $n\to \infty$ (where $g$ is the Yang--Mills
coupling constant)
leads to 't Hooft's topological expansion of Yang--Mills
theory in a power
series in $1/n$ \REF\thoofta{G.  't Hooft, {\it A planar
diagram theory for strong interactions}, Nucl. Phys. {\bf
B72} (1974) 461.}[\thoofta].  The diagrams in this expansion
correspond to the world-sheets
of string theory where the closed-string coupling constant,
$\kappa$, is related to the Yang--Mills coupling by
$\kappa = \alpha g^2$ (where $\alpha$ is a constant) and is
therefore
inversely proportional to $n$.  According to the folklore of
the large-$n$
approximation glueball states may be
associated with closed strings and meson states with open
strings carrying
quark flavour quantum numbers at their end-points, whereas
baryons are very massive solitonic states
\REF\wittena{E. Witten, {\it Baryons in the $1/N$
expansion}, Nucl.  Phys.
{\bf B160} (1979) 57.}[\wittena] (about which nothing will
be said in this paper).

Obviously a string theory of hadrons has to have rather
different properties from a string theory of the \lq
conventional' type
that describes gravity together with other forces.  Firstly,
the observed
spectrum of hadrons does not include
massless spin-1 and spin-2 gauge particles.  Furthermore,
fixed-angle
scattering cross sections decrease exponentially with energy
in conventional string theories whereas experimental
hadronic cross sections decrease as a power of the energy at
fixed angle --- a
crucial indication of point-like substructure.   An added
constraint is that
hadrons must couple correctly to external (electro-weak)
sources, which is a
natural property of QCD but does not seem possible for
string theories
describing gravity.

In order to compare QCD with string theory \REF\polchinew{J.
Polchinski,
{\it High temperature limit of the confining phase},
Austin preprint UTTG-26-91.}Polchinski recently [\polchinew]
considered the high temperature limit of the confining phase
of large-$n$ $SU(n)$ Yang--Mills theory (in this limit the
system is unstable since there is a deconfining phase
transition at a finite
temperature).   He discussed the correlation functions of
Wilson loops that
wrap $q$ times around the periodic imaginary time dimension
(where the period $\beta$ is the inverse temperature) and
are fixed at spatial points.  The
form of the one-loop free energy indicates that the
dominant states that mediate these gauge-invariant
correlation functions are tachyonic glueball states with
masses
$$M_{YM}^2 = - {2g^2 n\over \pi^2\beta^2 q^2}\eqn\masspect$$
as $\beta\to 0$ and to lowest order in perturbation theory.

The temperature-dependent spectrum of conventional tree-
level string
theory looks nothing like \masspect.  The \lq mass' spectrum
of
a closed-string theory with a compact euclidean time
coordinate of circumference $\beta$ is given (at tree level)
by
$$ \alpha' M_{string}^2 \equiv -  \alpha'{\bf p}^2 - {4\pi^2
\alpha'
m^2\over \beta^2} = {\beta^2 q^2 \over 4\pi^2 \alpha' } + 2N
+ 2\tilde N -4 ,\eqn\massclose$$
where ${\bf p}$ denotes the spatial components of the
momentum  $p^I$ ($I=1,\dots,D$) and $m$ is the integer
Kaluza--Klein charge of
the quantized $p^0$ (and $\alpha' =1/\pi T$, where $T$ is
the string tension).
At low temperature (large $\beta$) states
of non-zero winding have positive $M_{string}^2$.  The
Hagedorn transition at $\beta^2 = 16\pi^2 \alpha'$  is seen
to occur at the temperature at which the lowest-mass winding
state ($m=N=\tilde N=0$ and $q=1$) becomes massless
\REF\satha{B. Sathapalian, {\it Vortices on the string
world-sheet and
constraints on toral compactification}, Phys. Rev. {\bf D35}
(1987)
3277.}\REF\kogana{Ya.I. Kogan, {\it Vortices on the world-
sheet of a string;
critical dynamics}, JETP Lett. {\bf 45} (1987)
709.}\REF\wittena{J. Atick and E. Witten, {\it The Hagedorn
transition and the
number of degrees of freedom of string theory}, Nucl. Phys.
{\bf B310} (1988) 291.}[\satha-\wittena].  If the expression
for
$M_{string}^2$ is continued to $\beta \sim 0$ it is quite
dissimlar from
$\masspect$ in its dependence on $\beta$.  Accordingly,
Polchinski
suggested that a  string theory of QCD  would  be one in
which the effective number of world-sheet fields grows with
temperature at high temperature (in which case the $-4$ on
the right-hand side of \massclose\ is replaced by a
$\beta$-dependent term).
This paper will indicate that the temperature dependence of
string theory may be  substantially affected by the presence
of world-sheet boundaries and this may lead to an
(apparently) different resolution of the discrepency between
the
high temperature mass spectra of Yang--Mills theory and
string theory.

The fact that $M^2_{YM}$ is proportional to $g^2n$
means that it is associated with the lowest-order (genus
zero) diagrams in the topological expansion of the large-$n$
theory. It is usually assumed that such diagrams are related
to the tree diagrams of any corresponding string theory
since the string coupling constant $\kappa$ is proportional
to $1/n$.  For example, the radiative corrections
associated with a loop of closed string give corrections to
the mass spectrum of order $\kappa^2 \sim 1/n^2$ (and in any
case are not of the correct form to produce the Yang--Mills
mass spectrum \masspect\ at high temperature).  However,
more interesting corrections arise in presence of world-
sheet
boundaries.  The insertion of a single boundary (which
changes the genus of
the surface by $1/2$) is proportional to $w
\kappa$, where $w$ is a weight which may be associated with
internal symmetry in the Chan--Paton manner.  In the usual
formulation of open string theory the coordinate $X^\mu
(\sigma,\tau)$ satisfies Neumann boundary conditions, \ie,
the
normal derivative vanishes, $\partial_n X|_B =0$.  In this
case the theory describes interacting open and closed
strings, and the boundaries are identified with the
trajectories of free open-string end-points.  The original
string picture of
mesons associated quark quantum numbers of a $U(m)$ flavour
group with the
free endpoints so that a closed world-sheet boundary carries
the flavour weight
$w \equiv w_N =const. \times m$.   The value of the constant
is fixed by
imposing perturbative unitarity in open-string channels.
A string  theory defined by summing over Riemann surfaces
of arbitrary topology with these boundary conditions, which
will be
called a \lq Neumann' theory, describes interacting open and
closed strings.   An alternative choice of boundary
conditions requires the tangential derivative of the
coordinate to vanish at the boundary, $\partial_t X^\mu|_B
=0.$ \REF\greena{M.B. Green,
{\it Space-time duality and Dirichlet string theory},
Phys.  Lett. {\bf 266B} (1991) 325.}\REF\greenb{M.B.  Green,
{\it Point-like
structure in string theory}, Invited talk
at the first international Sakharov conference, Moscow, June
1991 (QMW-91-11, to be published).}  This constant Dirichlet
condition leads to a \lq
Dirichlet' theory, in which each world-sheet boundary is
fixed at a space-time point, $y_B^\mu$, thereby defining a
set of space-time correlation functions for point-like
closed strings.  Summing over all topologies and integrating
over the $y_B^\mu$'s gives a radically modified
Minkowski-space purely closed-string theory
([\greena,\greenb] and references therein).  The constant
Dirichlet  conditions result in power-behaved fixed-angle
scattering amplitudes.  The weight of a Dirichlet boundary
may be naturally associated with the colour group, in which
case $w\equiv
w_D = const. \times n$.  In that case $w_D \kappa \sim g^2
n$ and the
boundary insertion contributes at the same order as the tree
diagrams, even though it is of different genus.

The normalization
of $w_D$, including its sign, is not fixed by conventional
unitarity arguments.  It is conceivable that it may be
determined by
a more subtle \lq dual' form of unitarity  based on the fact
that the
Dirichlet theory can be related to the Neumann theory by
a spatial duality transformation in which all dimensions are
compactified on circles of radius $R$ in the limit $R\to 0$
[\greena].
The $y_B^\mu$ enter the compactified Neumann theory as
$U(1)$ gauge fields that
are constant on each boundary.  They may take different
values on different
boundaries in which case they must be integrated over their
allowed range.  [Alternatively the
$y_B^\mu$ can be considered to be boundary values of world
sheet scalar fields $Y^\mu(\sigma,\tau)$ (satisfying
$\partial_t Y^\mu|_B =0$) that couple to $\hat X^\mu$
(where  $\hat X$ denotes a coordinate satisfying Neumann
conditions) by a total derivative term proportional to
$\int d\sigma d\tau\epsilon^{\alpha\beta} \partial_\alpha
\hat X^\mu \partial_\beta Y_\mu$ in the action].
\FIG\fone{}
\figure{2.17in}
 \smallskip Figure~\fone.\
The insertion of a single boundary (indicated by thick
lines) in a closed-string
propagator.  (a)  A representation in which the world-sheet
exhibits intermediate open strings.  In the Dirichlet
theory these have end-points that are fixed at $y^\mu$ in
space-time, which is integrated over in the amplitude.  (b)
A representation that exhibits a closed-string tadpole.  (c)
In the light-cone gauge (in which $\tau=X^+$ and
$P^+(\sigma)=p^+/2\pi$) the Dirichlet boundary (represented
by a vertical slit) forces a finite fraction of the total
$p^+$ carried by the string to accumulate at a single
spatial point at \lq time' $\tau=X_0^+$.
\endfigure
Figure \fone\ illustrates three representations of the
insertion of a single boundary in a closed-string
propagator.  The first of these (fig \fone(a)) shows
the intermediate (group singlet) open strings that mix with
the closed ones.
In the case of the Neumann theory at zero temperature this
is the familiar mixing of open-string theory that leads, for
example, to a Higgs-type mechanism that gives a mass to the
level-one antisymmetric tensor state \REF\cremmera{E.
Cremmer and J. Scherk, {\it Spontaneous dynamical breaking
of gauge symmetry in dual models}, Nucl.  Phys.  {\bf B72}
(1974) 117.}\REF\kalba{M. Kalb and P. Ramond, {\it Classical
direct interstring action}, Phys.  Rev. {\bf D9} (1974)
2273.}\REF\shapirod{J.  Shapiro and C.B. Thorn,{\it BRST-
invariant
transitions between closed and open strings},
Phys. Rev. {\bf D36} (1987) 432.} [\cremmera-\shapirod]. In
the case
of the Dirichlet theory there are no physical
intermediate open-string states (apart from the level-1
vector which gives rise to a divergence in the
zero-temperature theory).  Furthermore, the position of the
boundary $y^\mu$ is to be integrated.  Figure \fone(b)
represents
the insertion as a closed-string tadpole coupling to the
vacuum via either a Neumann or a Dirichlet boundary.  The
presence of a massless dilaton in the cylinder channel makes
the diagram divergent
as usual.  In figure \fone(c) the Dirichlet boundary is
illustrated in the light-cone gauge.  It imposes an
instantaneous condition
(at light-cone time $\tau =X_0^+$) that forces a finite
fraction of the
light-cone momentum, $p^+$, to accumulate at a single
space-time point which is then integrated.

The difference between the temperature-dependence of the
expressions for the boundary
insertions in the Neumann and Dirichlet theories is quite
striking.  In the case of the Neumann theory there are
infinite numbers of intermediate open-string states in fig
\fone(a) carrying
the same momentum as the closed strings.  If a dimension is
compactified winding number is not conserved --- the
intermediate open string does not possess a
winding number (the Neumann boundary in
fig \fone(b) can carry winding number into the vacuum).
Thus, the inverse propagator for a Neumann open string at
inverse temperature $\beta$ carrying momentum $ p^\mu \equiv
(p^I, 2\pi
m/\beta)$ is given by
$$\Delta^{-1}_N = \alpha' {\bf p}^2 + {4\pi^2 m^2 \over
\beta^2} + N_j-1,\eqn\neuprop$$
where $N_j$ is the level of an open-string
state with occupation numbers labelled by $j$ and the normal
ordering constant $-1$ arises from 24 transverse dimensions
(or it may be made up from world-sheet conformal matter and
ghost fields).

The only physical open-string state of the zero-temperature
Minkowski-space
Dirichlet theory is an isolated level-1 $U(n)$-singlet
vector.
The winding number associated with
a toroidal compactification is now conserved --- the fixed
end-point intermediate open string carries the winding
number, $q$, of the closed strings to which it couples and
the propagator insertion is diagonal in winding number.  The
closed-string momentum  is not conserved if
$y$ is fixed but integration over $y$ imposes its
conservation and guarantees translation-invariance of the
Dirichlet theory.  The inverse propagator of an open
string with end-points fixed at $y_1$ and $y_2$ and inverse
temperature $\beta$
is given by [\greena]
$$\Delta_D^{-1} =  {1\over  4\alpha'\pi^2} (y_2-y_1 +
q\beta)^2
+ N_j -1. \eqn\dirprop$$
The intermediate open-string state in fig \fone(a) has
$y_1=y_2=y$ so that the propagator does not depend on the
$y_B$'s.  However, the open-string wave function still
depends on $y$  (\ie, there is a continuum of open-string
states and
momentum conservation follows upon integrating over $y$).

The matrix elements of the propagator insertion have the
form
$$\Pi_{rs} = \sum_j c_{r,j}\ \Delta_{open} \ c_{s,j}
,\eqn\neuminsert$$
where $\Delta_{open}$ represents either \neuprop\ or
\dirprop\
($\Delta_N$ and $\Delta_D$, respectively).  The coefficients
$c_{r,j}$
(which are proportional to the open-string coupling
constant, $g_{open}$) define the couplings
between a closed-string state with occupation numbers
labelled by $r$
and open-string states with occupation numbers $j$.  They
are functions of
$(\beta,  p^I, m, q)$ and include a group theory factor
$\delta^{ab}$
($a,b=1,\dots,n$) with indices that are contracted by the
propagator in
\neuminsert,  leading to an overall factor of $n$ in
$\Pi_{rs}$.
The shift in the closed-string masses is determined by
diagonalizing
$\Pi_{rs}$.  An important qualitative difference between the
temperature
dependence of the Neumann and Dirichlet theories can be seen
directly
from the
structure of \neuprop\ and \dirprop\ (with $y_1=y_2$).   In
the Neumann
case the propagator $\Delta_N$ has an unremarkable
dependence on $\beta$ in the limit $\beta \to 0$ (states
with $m\ne 0$ become infinitely massive and decouple,
leaving only $m=0$ states).  However, if the expression for
the Dirichlet propagator, $\Delta_D$,
is continued to $q\beta<< 2\pi \sqrt{\alpha'}$ (ignoring the
Hagedorn
transition at $\beta = 4\pi\sqrt{\alpha'}$) $\Pi_{rs}$ is
dominated by the
$N=1$, $m=0$ open-string states in \neuminsert.  These
open-string vector
states of non-zero winding have couplings $c_{r,1}$ to
closed-string states
labelled $r$. [The zero winding-number open-string state at
this level is the  vector Lagrange multiplier field that
causes
divergences in the
zero-temperature amplitudes in perturbation theory
[\greena].]  To first order
in $w_D\kappa$
this gives a shift proportional to $G_r w_D \kappa
/q^2\beta^2$ in the
$($mass$)^2$ of closed-string states
(recalling that $\kappa$ is proportional to $g_{open}^2$).
The coefficient
$G_r$ depends on precise details of the couplings,
$c_{r,1}$,  that will
be discussed  below, but is a smooth non-vanishing
function of $\beta$ for small $\beta$.
Therefore, when continued to higher temperatures, $\beta^2
<<\alpha'
G_r w_D \kappa/q^2 $ the shifted mass (at first order in
$w_D\kappa$)
may have the same general form as that in \masspect\ if the
weight of the boundary is chosen so that $w_D \sim n$ (as
expected for a
Chan--Paton factor associated with the colour $U(n)$).
Such a
comparison is purely formal in the absence of any estimate
of higher-order
contributions to both the QCD and the string expressions
(which will not be
considered in this paper).

\FIG\ftwo{}
\figure{2.17in}
 \smallskip Figure~\ftwo.\
The insertion of arbitrary numbers of boundaries in an
open-string propagator leads to higher-order mass
corrections.
\endfigure
The complete modification to the bare closed-string
propagator
due to boundary insertions, illustrated by
fig \ftwo, involves mass insertions of arbitrarily high
order in
$w\kappa$ as well as being non-diagonal in the
closed-string states.  However, to lowest order in $w
\kappa$
a single boundary insertion dominates $\Pi_{rs}$ and only
those
matrix elements that couple closed-string states of  equal
bare mass contribute.  These perturbative mass shifts are
then
determined by the on-shell amplitude for coupling the two
closed-string
states, $r$ and $s$, to a disk with appropriate boundary
conditions.    We shall first use this to consider the
simplest
case which arises when $r$ and $s$ are the (non-degenerate)
closed-string
tachyonic ground-states with non-zero winding number and
momentum
satisfying (from hereon the tension will be chosen so that
$\alpha'=1$ for convenience)
$$M_{tachyon}^2 = - {\bf p}^2  - {4\pi^2 m^2\over \beta^2} =
{q^2
\beta^2\over 4\pi^2} -4 .
\eqn\masstach$$

The two-tachyon amplitudes may be evaluated in the standard
manner (or extracted from eq.(14) of [\greena]).  With
Neumann conditions at
the boundary of the disk the amplitude with two ground
states is
$$\eqalign{\Pi_N=& w_N\kappa\int_0^1 {dx\over x^2}  (1-
x)^{{2\pi^2 m^2\over
\beta^2} - {q^2\beta^2\over 8\pi^2} + {{\bf p}^2\over 2}}
\cr
=& w_N\kappa\int_0^1 {dx\over x^2} (1-x)^{2 -
{q^2\beta^2\over 4\pi^2}},\cr}\eqn\neumtach$$
where the mass-shell condition \masstach\ has been used and
an overall constant has been arbitrarily absorbed into the
coupling since the normalizatiotion of the boundary
insertion is arbitrary
at this stage.  The $x=0$ boundary of moduli space (where
the two vertex
operators coincide) gives rise to singularities associated
with the  emission of a closed-string
tachyon and dilaton into the vacuum with zero momentum.
These
familiar divergences do not affect the dominant $\beta$
dependence at small $\beta$.  The boundary $x=1$ (the region
in which one of the vertex operators on the disk collides
with the boundary if
the other is held fixed at the centre of the disk) gives
singularities at $\beta^2= 4\pi^2(3+N)/q^2$ (integer $N$).
These
are the values of $\beta$ at which the intermediate open-
string
states are on shell and at which the mass corrections
are singular (the expression for $\Delta_N$,
\neuprop, has poles at these positions when \masstach\ is
used).  In the limit $\beta\to 0$ (which is non-singular for
finite $q$) the amplitude is constant.

With Dirichlet conditions the two-tachyon disk amplitude
becomes
$$\eqalign{\Pi_D=& \lim_{k\to 0} w_D\kappa \int_0^1 {dx\over
x^2} x^{k^2\over 4} (1-x)^{-{2\pi^2 m\over
\beta^2} + {q^2\beta^2\over 8\pi^2} -{{\bf p}^2\over 2}} \cr
=& \lim_{k\to 0} w_D\kappa \int_0^1 {dx\over x^2}
x^{k^2\over 4}
(1-x)^{-2+{q^2\beta^2\over 4\pi^2}}\cr=& \lim_{k\to 0}
w_D\kappa {\Gamma(-
1+{q^2\beta^2\over 4\pi^2})\Gamma(-1+{k^2\over 4})
\over \Gamma (-2 + {q^2\beta^2\over 4\pi^2} + {k^2\over 4})}
,\cr}\eqn\diritach$$
where $k^\mu = - p^{1\mu} - p^{2\mu}$ is the momentum that
leaks through the Dirichlet boundary but is set to zero by
the $y^\mu$
integration (and $p^\mu \equiv p^{1\mu}$).
Again there is a $x=0$ singularity due to the emission of a
dilaton at zero momentum (the Dirichlet condition avoids the
zero-momentum tachyon
divergence by defining $k^2=0$ by analytic continuation from
$k^2>4$),
which does not affect the dominant $\beta$ dependence at
small $\beta$.  The $x=1$ boundary gives rise to
singularities at $q^2\beta/4\pi^2 = 1-N$ (which are the
positions of the poles of $\Delta_D$ given by \dirprop) that
are
contained in the factor $\Gamma(- 1+q^2\beta^2/4\pi^2)$.
For $q\beta <<2\pi $ the mass shift behaves as $8\pi^2
w_D\kappa
/q^2\beta^2$ in accord with the earlier intuition.
Furthermore,
since the modified closed-string propagator has the form
$(p^2 +
m^2  - \Pi)^{-1} \sim (p^2 + m^2)^{-1}\ \Pi (p^2 + m^2)^{-1}
+ \dots$ the mass
correction is negative if $w_D$ is chosen to have positive
sign.
The above argument assumed the shift is small relative to
the tachyon mass,
\ie, $4 >>8\pi^2 |w_D| \kappa/q^2 \beta^2 $.
If the result is nevertheless continued
to values of $\beta \sim 0$ so that $M_{tachyon}^2 + \delta
M_{tachyon}^2
\sim \delta M^2_{tachyon}$ the leading-order tachyon
spectrum  formally
resembles that of \masspect, as was anticipated earlier.
Tree-level corrections to the tachyon mass
of higher order in $w\kappa$ arise from the insertion
of more boundaries in the disk (for example, the diagram
with two boundaries
is an annulus that can be described as a loop of open
string).

We now turn to consider the shift in the masses of arbitrary
closed-string states to leading order in $w_D
\kappa$.  These are determined by the operator, $\Pi$, that
inserts a single Dirichlet boundary in the world-sheet.  The
propagator joining two arbitrary
closed-string states, $\langle X|$ and
$| Y\rangle$, of momenta  $p^\mu \equiv (p^I,m)$ and winding
number $q$, can be formally expressed as a sum,
$$\eqalign{G(X,Y) = &\langle X|\Delta_C + \Delta_C \Pi
\Delta_C + \dots | Y\rangle \cr
              =&  \langle X|\Delta_C (1-\Pi\Delta_C)^{-1}|
Y\rangle \cr} \eqn\fullprop$$
where the bare closed-string propagator, $\Delta_C$, is
given by
$\Delta_C^{-1} =  p^2 +\beta^2 q^2 / 4\pi^2  + 2N
+ 2\tilde N -4$ with $N=\tilde N$ (and $p^2 =  {\bf p}^2 +
4\pi^2
m^2/\beta^2$).  The boundary insertion
operator, $\Pi$, has the form
$$ \Pi = w_D\kappa\Upsilon^\dagger \Delta_D \Upsilon,
\eqn\closedopen$$
where $\Upsilon$ is the operator that couples a closed
string to an open string  with fixed end-points (in the
winding-number $q$ sector) and $\Delta_D$ is given by
\dirprop.

For $q\beta <<2\pi$, $\Pi$ is dominated by the pole of
$\Delta_D$ due to the
 $N=1$ open-string vector state with winding-number $q$.
This state couples to a Dirichlet boundary via the
interaction $\Upsilon\sim
p^\mu |B\rangle$\foot{This is the coupling between the
isolated physical
open-string state and the closed-string states that couple
to the boundary.
More generally, unphysical string states contribute to non-
leading effects.},
where $|B\rangle$ is the boundary state defined to satisfy
the
Dirichlet condition, $\partial_\sigma X^\mu (\sigma)
|B\rangle =0$.
The contribution to $\Pi$ of this state is then given by
$$\Pi = K |B\rangle \langle B|, \eqn\noneins$$
where $K = 2\pi^2 w_D \kappa p^2 / q^2 \beta^2$.  It
therefore
contributes to the shift in mass of all the scalar closed-
string states.
For example, consider the finite set of equal-mass
Fock-space end-states (labelled $r$ and $s$) at a given
level
with bare mass ($M^2_r= M_s^2 =4N_r -4 + \beta^2 q^2
/4\pi^2$).  The shift in the position of the poles of
\fullprop\ can be
estimated to first order in $w \kappa$ by considering
momenta such that $p^2 \sim -M^2_r$ in \fullprop.
Substituting this and \noneins\ into \fullprop\ gives
$$\eqalign{G(r,s) \sim & {1\over p^2 + M_r^2}  \left[
\delta_{rs} +
{ K B_r B_s \over p^2 + M_r^2 - K B^2}   \right]   \cr
     \sim &    {1\over p^2 + M_r^2}\left(\delta_{rs} - {B_r
B_s \over B^2}\right) + {1\over p^2 + M^2_r -  K  B^2}
 \left({B_r B_s \over  B^2}\right) +  non{\rm -}pole\ terms,
\cr}\eqn\poleshift$$
where $B_r = \langle r| B\rangle$, $B^2 \equiv \sum_r B_r
B_r$ and $KB^2$ has been assumed to be small.  If the level
is non-degenerate (\ie, $r$ labels a
single state) the first term in the second line of
\poleshift\ (the \lq transverse' term) vanishes and the the
pole position is shifted to $M_r^2 + \delta M_r^2$, where
$\delta M_r^2 = -KB^2 \sim  2\pi^2 w_D \kappa M_r^2 / q^2
\beta^2$
(where the substitution $p^2 =- M_r^2$ has been made in
$K$, which is valid for calculations to lowest order in
$w_D\kappa$).   More generally, the levels are highly
degenerate.  In such cases only one combination of states in
\poleshift\  (the \lq longitudinal one', $|s\rangle
B_rB_s/B^2$) is shifted --- again $\delta M^2_r =- K  B^2$.
The
remaining states in \poleshift\ remain unshifted.
If $w_D$ is chosen to be positive the shifts $\delta M_r^2$
are
positive for states of positive $M_r^2$, the massless state
(the dilaton) remains massless and the tachyon develops a
more negative value of (mass)$^2$ of the form determined
earlier.
If $w_D$ is
negative the signs of the shifts are reversed and the
possible correspondence
with \masspect\ is lost.

\REF\greens{M.B.  Green, {\it Wilson loops for strings and
superstrings at finite temperature}, QMW-91-19.}The tachyon
state of  winding number $q=1$ that plays an important
r\^ole in determining the leading small $\beta$ behaviour
when $w_D>0$ is
precisely the state invoked in [\satha-\wittena] as the
origin of the Hagedorn transition at $\beta= 4\pi$.   This
is not a physical state of the zero temperature
theory.  An analogous unphysical tachyon plays a similar
r\^ole in inducing the Hagedorn transition in superstring
theory, where there is no zero-temperature tachyon state
(and therefore the theory has a stable vacuum) [\wittena]. A
more consistent Dirichlet string theory than the one
considered in this paper based on the ordinary bosonic
theory
(perhaps some version of the superstring theory of
[\greens]) may
be free of physical tachyons but might still display the
$\beta\to 0$ behaviour characteristic of large-$n$ Yang--
Mills theory.

In summary, this paper has considered the effect of the
presence of world-sheet boundaries on the temperature
dependence of
bosonic string theories at high temperature.  In general,
this leads to a shift in the masses of states of non-zero
winding number
($q\ge 1$).
With Dirichlet boundary conditions this effect can be
particulary significant
at high temperature
since to first order in $w_D\kappa$ the mass shifts behave
as $1/\beta^2$ for   $q\beta <<2\pi $.  Formally (\ie,
ignoring the potentially
important effects of terms of higher order in perturbation
theory), the
$\beta \to 0$ limit of the theory with $w_D>0$ has a
tachyonic spectrum
analogous to that of  large-$n$ $U(n)$
Yang--Mills theory discussed by Polchinski [\polchinew].
Any such connection requires the weight,
$w_D$, associated with a Dirichlet boundary to be
proportional to $n$ so that
that the boundary insertions contribute at the same order as
the
usual string tree diagrams even though they are of higher
genus.     The precise relevance of this connection is
difficult to assess, particularly since the comparison is
between
the continuation to high temperature of the low-temperature
phases of the two theories --- the existence of the
deconfining phase
transition of Yang--Mills theory and the Hagedorn transition
of string theory
signifies that these phases are unstable and the spectrum is
tachyonic.
Furthermore, this comparison may be radically affected by
higher-order terms that contribute to both theories.
Nevertheless, together with the fact that the
Dirichlet theory possesses power-behaved fixed-angle
scattering amplitudes, this could be circumstantial evidence
of a
correspondence between Yang--Mills theory and some more
realistic version of
Dirichlet string theory.

The $q\ne 0$ terms
(terms of non-zero winding number) that dominate the
$\beta\to 0$ limit are infinitely suppressed as $\beta\to
\infty$.   The dominant ($q=0$) term of the zero-temperature
theory
has an apparently problematic divergence
due to the presence of  the $N=1$ open-string vector
intermediate state.  This  acts like a Lagrange multiplier
field [\greena] which generates a significant
shift of the bare closed-string spectrum which remains to be
elucidated.  Such a shift indicates the inadequacy
of string  perturbation theory --- the tree diagrams are
substantially
modified even for small $w_D \kappa$.
Arguments based on [\greena] suggest that
Dirichlet boundaries might lead to a  modification to the
usual gauge invariance
so that the perturbative graviton, open-string vector and
dilaton fields
($h^{\mu\nu}$, $A^\mu$ and $\phi$) transform (in linearized
approximation) as
$$\delta h^{\mu\nu} = \partial^{(\mu}\xi^{\nu)},
\qquad \delta A^{ab\mu} = g \xi^\mu \delta^{ab}
+ \partial^\mu \Lambda^{ab}, \qquad \delta \phi = g
\Lambda^a_{\ a},\eqn\trans$$
plus terms that vanish in the $\alpha'\to 0$ limit.  These
transformations are reminiscent of the transformations of
St\"uckelberg
fields that enter into  the gauge-invariant formulation of
the Pauli--Villars action for a massive
spin-two field of mass $\sim w_D\kappa$.  The absence of a
massless spin-two
state would be most welcome if
the theory is of relevance to hadronic physics but
calculations of low order
diagrams have not so far revealed the existence of such a
mass term explicitly.

All the considerations in this paper are based on standard
perturbation theory rules for critical string theory (in
which any number of spatial dimensions can be replaced by
conformal fields in a more or less standard manner).
It may be that these ideas
have more direct relevence  in the context of string theory
in sub-critical dimensions.
In any case the presence of important boundary  effects in
Dirichlet string theory that modify  tree-level amplitudes
suggests that the standard determination of the critical
dimension is likely to be significantly modified.

\refout

\bye